\documentclass[runningheads]{llncs}

\usepackage{times}

\usepackage[dvipsnames]{xcolor}
\usepackage{xcolor}
\usepackage{mdframed}

\usepackage{pgfplots}
\pgfplotsset{compat=1.17}
\usepackage{caption}
\usepackage{subcaption}

\usepackage{tikz}
\usetikzlibrary{shapes.geometric, arrows}

\usepackage{amsmath}
\usepackage{pifont} 
\usepackage{hyperref}
\usepackage{url}

\definecolor{boxcolor}{RGB}{142, 168, 229} 

\definecolor{boxcolor}{RGB}{204, 226, 255} 

\usepackage{graphicx} 
\usepackage{booktabs} 
\usepackage{tabularx}
\usepackage{wasysym}

\newcolumntype{M}[1]{>{\centering\arraybackslash}m{#1}}
\newcolumntype{L}[1]{>{\raggedright\arraybackslash}m{#1}}
\newcolumntype{C}[1]{>{\centering\arraybackslash}m{#1}}

\usepackage{multirow}
\usepackage{array}    
\usepackage{xspace}

\usepackage{color}
\usepackage{xcolor}
\newcommand{\BfPara}[1]{\vspace{0.3em}{\noindent\bf#1.}\xspace}

\newcommand{\etal}{{\em et al.}\xspace}

\begin{document}

\title{Simple Perturbations Subvert Ethereum Phishing Transactions Detection: An Empirical Analysis}
\author{Ahod Alghureid \and David Mohaisen}
\institute{University of Central Florida}
\maketitle

\begin{abstract} 

This paper explores the vulnerability of machine learning models, specifically Random Forest, Decision Tree, and K-Nearest Neighbors, to very simple single-feature adversarial attacks in the context of Ethereum fraudulent transaction detection. Through comprehensive experimentation, we investigate the impact of various adversarial attack strategies on model performance metrics, such as accuracy, precision, recall, and F1-score. Our findings, highlighting how prone those techniques are to simple attacks, are alarming, and the inconsistency in the attacks' effect on different algorithms promises ways for attack mitigation. We examine the effectiveness of different mitigation strategies, including adversarial training and enhanced feature selection, in enhancing model robustness. 

\end{abstract} 

\section{Introduction}\label{sec:introduction}
The proliferation of machine learning models in various domains has brought significant advancements in decision-making processes. However, concerns regarding the robustness and security of these models have also emerged alongside these advancements. Adversarial attacks, wherein small, carefully crafted perturbations are introduced into the input data to cause misclassification, seriously threaten the reliability of machine learning systems. Understanding the susceptibility of these models to adversarial attacks is crucial for developing robust and trustworthy AI systems.

The increasing prevalence of machine learning in cybersecurity applications has significantly improved the detection and prevention of various cyber threats. Among these threats, fraudulent activities such as phishing, scamming, and fake initial coin offerings (ICOs) pose substantial financial and personal data security risks. Machine learning models, particularly classification algorithms, have been deployed to identify and mitigate these threats with notable success. However, studies have raised concerns about the robustness and reliability of these models \cite{GoodfellowSS14,AbusnainaWAWWYM21,AlasmaryAJAANM20,AbusnainaJKNM20,AbusnainaKA0AM19,AbusnainaAAAJNM22,AbusnainaAAAJSN22}. The attacks presented in the literature (see section~\ref{sec:related}) are effective yet sophisticated. 

This study investigates the impact of extremely simple adversarial attacks on different machine learning models used in fraudulent transaction detection, specifically focusing on Random Forest (RF), Decision Tree (DT), and K-Nearest Neighbors (KNN) classifiers. By employing the Fast Gradient Sign Method, a widely recognized adversarial attack technique, we assess the performance degradation of these models when subjected to simple, realistic adversarially crafted inputs \cite{SzegedyZSBEGF13}. The primary objective is to understand how these models are susceptible to adversarial manipulation and explore potential mitigation strategies to enhance their robustness.

Previous research has highlighted the vulnerability of ML models to adversarial attacks \cite{PapernotM0JS16}. This study contributes to the existing body of knowledge by providing a detailed empirical analysis of the effects of simple adversarial attacks on fraud detection models and proposing practical approaches to mitigate these effects. Our findings reveal inconsistency across algorithms for their tolerance of simple manipulation, underscoring the importance of selecting appropriate models and implementing robust defense mechanisms tailored to specific applications.

\section{Literature Review}\label{sec:related}

Transaction fraud is common in cryptocurrency and calls for fraud detection. Agarwal~\etal~\cite{AgarwalTS22} analyzed malicious Ethereum transactions, including phishing and scams, to enhance threat detection capabilities. Rabieinejad~\etal~\cite{RabieinejadYPD23} employed Ethereum transaction data to develop models for identifying cyber threats. Sanjalawe~\etal~\cite{SanjalaweA23} used the Benchmark Labeled Transactions Ethereum dataset to detect abnormal transactions related to illicit activities. Zola~\etal~\cite{ZolaSBGU22} analyzed WalletExplorer data to classify Bitcoin addresses and reduce anonymity, aiding in identifying entities involved in illegal activities.  Yang~\etal~\cite{YangLLWGZ20} focused on applying transaction datasets in the Bitcoin network to detect spam transaction attacks. Mozo~\etal~\cite{MozoPPCT21} applied transaction datasets to detect cryptomining attacks within the Monero

Recent research demonstrated the vulnerability of deep convolutional neural networks (CNNs) to adversarial attacks~\cite{NarodytskaK17}. In~\cite{CartellaAFYAE21}, Cartella et al. adapted algorithms to imbalanced tabular data for fraud detection, achieving a perfect attack success rate. Bhagoji~\etal~\cite{BhagojiHLS18} introduced a Gradient Estimation black-box attacks to a target model's class probabilities, achieving near-perfect success rates for both targeted and untargeted attacks on deep neural networks (DNNs).

Generative Adversarial Networks (GANs) generate adversarial examples (AEs) and handle data perturbations in cryptocurrency transaction datasets. Table~\ref{table:RW} summarizes these methods and highlights key findings across different domains, underscoring the diverse methods employed to address security challenges in cryptocurrency networks. For instance, Fidalgo~\etal~\cite{FidalgoCP22} utilized GANs for synthetic data generation and augmentation to tackle class imbalance in Bitcoin. Agarwal~\etal~\cite{AgarwalTS22} employed Conditional GANs (CTGAN) to create realistic adversarial data for Ethereum transactions.

Rabieinejad~\etal~\cite{RabieinejadYPD23} used Conditional GANs (CTGAN) and Wasserstein GANs (WGAN) to generate synthetic samples for augmenting Ethereum transaction datasets, enhancing detection capabilities against cyber threats. Sanjalawe~\etal~\cite{SanjalaweA23} used Semi-Supervised GANs and feature extraction techniques to perturb features in Ethereum transaction datasets. Zola~\etal~\cite{ZolaBBGU20} focused on data augmentation using various GAN configurations to generate additional data for underrepresented classes in Bitcoin transactions. Zola~\etal~\cite{ZolaSBGU22} investigated different GAN architectures to generate synthetic Bitcoin address data to mitigate class imbalance and improve entity classification. 

Yang~\etal~\cite{YangLLWGZ20} used a Wasserstein Generative Adversarial Network with divergence (WGAN-div) to generate adversarial examples for spam transaction detection in the Bitcoin network. Mozo~\etal~\cite{MozoPPCT21} utilized WGANs to create synthetic network traffic data to detect cryptomining attacks in the Monero (XMR) network.

\begin{table}[ht]
\centering
\caption{Summary of some of the prior work.}\label{table:RW}
\scalebox{0.8}{\begin{tabular}{|l|c|l|l|}
\hline
\textbf{Paper Title} & \textbf{Year} & \textbf{Adversarial Techniques} & \textbf{Applications in Cryptocurrency} \\
\hline
Li~\etal~\cite{LiCGN18} & 2018 & GANs & Anomaly detection, Secure Water Treatment \\
\hline
Qingyu~\etal~\cite{GuoLAHHZZ19} & 2019 & IFCM, AIS, R3 & Fraud detection, TaoBao \\
\hline
Ba~\etal~\cite{Ba19} & 2019 & GANs & Credit card fraud, 31-feature dataset \\
\hline
Ngo~\etal~\cite{NgoWKPAL19} & 2019 & GANs & Anomaly detection, MNIST, CIFAR10 \\
\hline
Zola~\etal~\cite{ZolaBBGU20} & 2020 & GANs, data augmentation & Bitcoin, WalletExplorer (categorized addresses) \\
\hline
Yang~\etal~\cite{YangLLWGZ20} & 2020 & WGAN-div, GRU-based detection & Bitcoin, custom spam transaction dataset \\
\hline
Shu~\etal~\cite{ShuLKT20} & 2020 & GANs & Intrusion detection, network traffic \\
\hline
Mozo~\etal~\cite{MozoPPCT21} & 2021 & WGANs, synthetic traffic & Monero, custom cryptomining dataset \\
\hline
Fursov~\etal~\cite{FursovMKKRGBK0B21} & 2021 & Black-box attacks & Transaction records \\
\hline
Agarwal~\etal~\cite{AgarwalTS22} & 2022 & CTGAN, K-Means Clustering & Ethereum, Etherscan dataset (2,946 malicious accounts) \\
\hline
Fidalgo~\etal~\cite{FidalgoCP22} & 2022 & GANs, data augmentation & Bitcoin, Elliptic dataset (200K+ transactions) \\
\hline
Zola~\etal~\cite{ZolaSBGU22} & 2022 & Various GANs, adversarial learning & Bitcoin, WalletExplorer (16M+ addresses) \\
\hline
Rabieinejad~\etal~\cite{RabieinejadYPD23} & 2023 & CTGAN, WGAN & Ethereum, 57K normal, 14K abnormal transactions \\
\hline
Sanjalawe~\etal~\cite{SanjalaweA23} & 2023 & GANs, feature extraction & Ethereum, labeled transactions (normal, abnormal) \\
\hline
\end{tabular}}
\end{table}

\BfPara{Research Gap}Despite advancements in machine learning and blockchain technology, critical gaps persist in effectively understanding mitigating adversarial attacks and emerging security threats within cryptocurrency networks. AEs grounded in the context of application are underexplored. AEs in the feature space that leverage targeted and untargeted attacks, transaction fraud, smart contract exploits, etc., are lacking. This underscores the imperative for further analysis. This study aims to bridge this gap by investigating the robustness of machine learning-based phishing detection algorithms against {\em simple manipulations}, comparing the effectiveness of various algorithms in resisting such attacks, and exploring mitigation strategies to enhance their resilience. Our approach involves evaluating the algorithms' susceptibility to subtle feature manipulations and conducting a comparative analysis to identify the most robust models.

\section{Research Questions}\label{RQ}
This research explores the robustness, comparative performance, and mitigation strategies of machine learning-based phishing detection algorithms for Ethereum transactions in adversarial manipulations. The following questions highlight the core issues and the necessity to address them in light of existing literature.

\noindent {\bf RQ1.}{\bf Are machine learning-based phishing detection algorithms for Ethereum robust against simple manipulations of individual features?}
This question is motivated by the vulnerability of machine learning (ML) models to adversarial attacks, as demonstrated in several studies~\cite{NarodytskaK17,CartellaAFYAE21,BhagojiHLS18,SilvaN20,ChenZBH19}. In these studies, slight modifications in input data significantly altered the classification outcome, highlighting how even minor changes in key features like transaction amounts or timestamps can lead to incorrect classifications. Such manipulations can make it easier for adversaries to deceive models, underscoring the need to evaluate the robustness of ML models used in Ethereum phishing detection to ensure they remain effective and reliable despite such attacks.

\noindent{\bf RQ2.}{\bf How do different machine learning algorithms compare robustness against adversarial manipulations in Ethereum phishing detection?}
This question stems from the observation that various machine learning algorithms, while effective in detecting phishing activities, exhibit differing levels of resilience against adversarial attacks~\cite{SinghH19,OliveiraVSVBVG21,FidalgoCP22,AgarwalTS22,RabieinejadYPD23,CroceASDFCM021}. The robustness of these algorithms can vary significantly under adversarial conditions, which can be seen in studies where some algorithms perform better than others when subjected to adversarial manipulations designed to evade detection~\cite{LiCGN18,DingWZYFG19,Stutz0S19}. A comparative analysis of these algorithms is essential to identify those that provide the best defense against such threats, ensuring the highest possible security in Ethereum transaction classification.

\noindent{\bf RQ3.} {\bf How can the impact of manipulations be mitigated in machine learning-based Ethereum phishing detection?}
This question arises from the need to develop robust defensive strategies against adversarial attacks, as highlighted in recent literature~\cite{YangLLWGZ20,MozoPPCT21,CarmonRSDL19,CohenRK19,XieWMYH19}. Effective mitigation strategies could include enhancing feature selection processes, employing advanced data augmentation techniques, or implementing adversarial training methods~\cite{LiCH19,RabieinejadYPD23,ZolaBBGU20}. Understanding and developing these techniques are crucial for improving the security and reliability of ML-based phishing detection systems in Ethereum transactions, thereby reducing the risk posed by adversaries manipulating transaction data and enhancing overall network security.

\section{Methodology}\label{sec:methodolgy}

\begin{figure}[t]
  \centering
  \includegraphics[width=8.5cm]{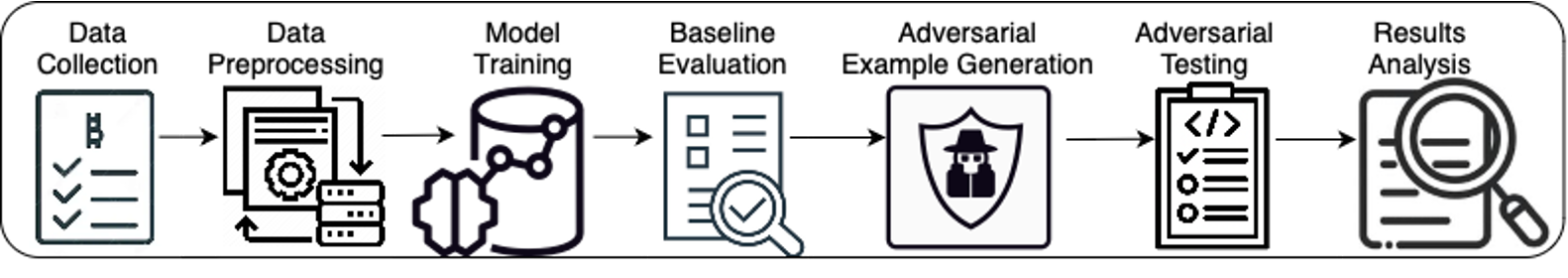}\vspace{-2mm}
  \caption{Pipeline in Ethereum transactions and adversarial testing.}
  \label{fig:method}\vspace{-3mm}
\end{figure}

Our pipeline is shown in \autoref{fig:method} and some of its key aspects are reviewed below.

\subsection{Data Preparation}

\subsubsection{Dataset Description}

For our analysis, we employed two datasets. The first dataset, as detailed by Kabla~\etal~\cite{KablaAMK22}, focuses on binary classification, distinguishing between phishing and benign transactions. This dataset encompasses features such as \texttt{TxHash}, a unique identifier for each transaction; \texttt{BlockHeight}, which specifies the height of the block in which the transaction was recorded; and \texttt{TimeStamp}, indicating the exact time the transaction was confirmed and added to the blockchain. It also includes the \texttt{From} and \texttt{To} addresses, representing the Ethereum addresses of the sender and receiver, respectively. The \texttt{Value} feature denotes the amount of Ether transferred in the transaction, while \texttt{ContractAddress} indicates the Ethereum address of the smart contract involved, if applicable. Additionally, the \texttt{Input} field contains any extra data provided with the transaction. The dataset labels transactions as either phishing (1) or benign (0) under the \texttt{Class} feature. Dataset 1 includes 23,472 transactions, of which 15,989 are benign and 7,483 are phishing.

The second dataset, as described by Al-Emari~\etal~\cite{Al-EmariASM20}, is intended for multi-class classification, categorizing transactions into Phishing, Scam, Fake ICO, or Benign. This dataset features the \texttt{hash}, a unique identifier for each transaction, and \texttt{nonce}, a counter ensuring each transaction is processed only once. The \texttt{transaction\_index} indicates the transaction's position within the block, while \texttt{from\_address} and \texttt{to\_address} denote the blockchain addresses of the sender and receiver, respectively. The \texttt{value} field specifies the amount of cryptocurrency transferred. The dataset also includes \texttt{gas}, representing the gas limit provided for the transaction, and \texttt{gas\_price}, indicating the price per gas unit. The \texttt{input} field contains additional data attached to the transaction. Moreover, \texttt{receipt\_cumulative\_gas\_used} provides the total gas used by all transactions up to and including the current one within the block, and \texttt{receipt\_gas\_used} specifies the gas consumed by this particular transaction. The \texttt{block\_timestamp} and \texttt{block\_number} detail the time and number of the block that includes the transaction, while \texttt{block\_hash} serves as the block's unique identifier. Lastly, the \texttt{from\_scam} and \texttt{to\_scam} fields indicate whether the sender's and receiver's addresses are associated with scams (0 for no, 1 for yes). The dataset also includes categorical data, \texttt{from\_category} and \texttt{to\_category}, which classify the nature of the participants, such as Phishing, Scamming, or Fake ICO. In this dataset, \texttt{Benign} had 57,000 transactions (79.28\%),  \texttt{Scamming} had 11,143 transactions (15.51\%), \texttt{Phishing} had3,106 transactions (4.32\%), and \texttt{Fake ICO} had only 1 transaction. Dataset 2 includes 71,250 transactions, with 80-20 training-testing splits.

\subsection{Experimental Procedures}

We utilized the two datasets outlined earlier to examine how simple AEs impact the classification accuracy of three classifiers: RF, DT, and KNN.

\BfPara{Minimal Manipulations} AEs were crafted by manipulating specific features. For the first dataset, we manipulated for feature. \ding{172} {\em Timestamp Manipulation (\textit{TimeStamp})}: We use predefined intervals to simulate future occurrences, testing the models' robustness to shifts against hypothetical scenarios of temporal manipulation. \ding{173} {\em Value Manipulation (\textit{Value})}: Transaction values were altered using two strategies: uniformly by adding a fixed percentage to each transaction's value or proportionally by introducing a random percentage change relative to each transaction's original value. \ding{174} {\em Receiver Address Manipulation (\textit{To})}: The receiver's address was randomly changed to different addresses within the dataset to simulate phishing transactions directed to alternative destinations. \ding{175} {\em Sender Address Manipulation (\textit{From})}: The sender's address was altered to different sender addresses to simulate phishing transactions originating from various sources. 

We implemented a two-pronged approach, including both targeted and untargeted adversarial attacks on a second dataset~\cite{Al-EmariASM20}. AEs were generated by focusing on a broader array of critical transaction features, allowing us to simulate realistic attack scenarios and identify potential vulnerabilities more precisely.

\BfPara{Untargeted Attacks} The study of untargeted attacks involved generating AEs by applying broad, random perturbations across the entire feature space. Initially, AEs were generated using all features to evaluate the model’s capacity to withstand attacks.

Individual features were then targeted to challenge the model more effectively. Modifying the \textit{from\_address} and \textit{to\_address} fields introduced new, unseen addresses to assess the model's ability to handle transaction origin and destination changes. Altering \textit{value}, \textit{gas} and \textit{gas\_price} simulated economic fluctuations, providing insights into the model’s sensitivity to variations in transaction costs. Manipulating \textit{block\_timestamp} and \textit{block\_number} mimicked transaction timing and sequence changes to understand the model's response to variations in transaction order. Altering  \textit{input}, \textit{receipt\_cumulative\_gas\_used}, and \textit{receipt\_gas\_used} help explore the impact of changes in content.

\BfPara{Targeted Attacks} We conducted targeted adversarial attacks focusing on three scenarios: benign, phishing, and scamming. We employed two methods for generating these targeted attacks: rule-based and gradient-based (using the Fast Gradient Sign Method).

\noindent{\em Rule-based Modifications} This approach applies straightforward, rule-based modifications to critical features like transaction value and timestamps, simulating realistic variations that can alter the classification outcomes. 

\begin{itemize}
    \item \textbf{Benign Targeted Attack:} This scenario aimed to assess the model’s ability to maintain a benign classification despite manipulations, simulating tactics used to camouflage malicious activities. We created artificial benign transactions by minorly adjusting features such as transaction value and block\_timestamp. 
    
    \item \textbf{Phishing Targeted Attack:} The focus was on modifying phishing-labeled transactions to evade detection by misclassifying them as benign. This involved altering attributes such as from\_address, to\_address, and value, simulating an adversary's attempt to bypass security measures. 

    \item \textbf{Scamming Targeted Attack:} In this scenario, scam-labeled transactions were manipulated to explore whether they could be misclassified as benign or other types. Adjustments to features like gas and gas\_price were made to examine the model against efforts to obscure scam activities through changes in transaction costs.
\end{itemize}

\paragraph{FGSM}
FGSM applied small, strategically calculated changes to subtly influence the model’s predictions while maintaining realistic feature values. Unlike the rule-based method, FGSM retained the existing distribution from the dataset, focusing on fine-tuning modifications based on the gradients to maximize the attack's effectiveness.

\begin{itemize}
    \item \textbf{FGSM Details:} FGSM calculates perturbations that align with the gradient direction of the loss function. The perturbations were applied using the formula:
    \[
    x' = x + \epsilon \cdot \text{sign}(\nabla_x J(\theta, x, y)), 
    \]

    where \( x \) is the original feature, \( x' \) is the perturbed feature (AE), \( \epsilon \) is a small scalar controlling the perturbation's magnitude, \( \nabla_x \) represents the gradient of the loss function \( J \) for \( x \), and \( J(\theta, x, y) \) is the loss function dependent on the model parameters \( \theta \), input \( x \), and true label \( y \). The term \(\text{sign}(\nabla_x J(\theta, x, y))\) provides the direction in which to perturb the feature vector to maximize the loss function.

    \item \textbf{Features:} FGSM perturbations were applied to transaction value, gas, gas\_price, and block\_timestamp. These features are critical in determining the nature of the transaction, and even small changes may affect the classification results. 

\end{itemize}

\section{Results and Analysis}

\subsection{Preliminary Results}
We evaluate how different perturbations on the first dataset affect the accuracy and resilience of RF, DT, and KNN models. We examine how uniform and proportional value manipulations and address change impact model performance. 

\BfPara{Timestamp Manipulations} 
We evaluate the effects of various timestamp modifications on classifier accuracy, including shifts of +24 hours, +1 hour, +30 minutes, +15 minutes, and +5 minutes. The RF classifier showed the highest resilience with only minor reductions in accuracy. For example, a one-day timestamp shift resulted in a decrease in accuracy from 98.82\% to 95\%, while a one-hour change led to an accuracy of 97.31\%. In contrast, the DT model experienced a more pronounced decline, with accuracy dropping from 98.35\% to 94.46\% with a one-day shift. The KNN classifier was most affected by these temporal manipulations, with accuracy falling from 94.45\% to 83.42\% for a one-day change. These findings highlight the superior robustness of the RF model in handling timestamp alterations, positioning it as a preferable option for phishing detection in environments with variable timestamps (Table \ref{table:time_minap}).

\BfPara{Value Manipulations} 
The uniform value manipulations caused significant declines. RF's accuracy dropped to 69\%, and DT's fell to 68\%, with substantial reductions in precision and recall for phishing transactions. Notably, the recall for phishing in DT decreased to 0.01\% under uniform manipulation. Proportional manipulations had minimal impact, with accuracy and other metrics remaining close to their original values. KNN maintained relatively stable performance across both manipulation types, indicating robustness against such value changes (Table \ref{table:value_manipulations}). 

\BfPara{Address Manipulations} The robustness of the classifiers was tested against AEs of the \texttt{From} and \texttt{To} address features, involving changes in 5,000, 10,000, and 23,472 instances. For the RF model, accuracy decreased to 87\% when the \texttt{From} address was manipulated and to 84\% for the \texttt{To} address. Precision and recall for phishing transactions also declined significantly, with F1-scores dropping notably. The DT model showed a moderate reduction in accuracy, dropping to 92\% for \texttt{From} and 93\% for \texttt{To} manipulations, and a noticeable decrease in recall for phishing. KNN was the most sensitive to these manipulations, with accuracy falling to 85\% for \texttt{From} and 93\% for \texttt{To}, and significant drops in precision and recall for phishing transactions (Tables \ref{table:from_result} and \ref{table:to_result}).

\begin{table}[htb]
\centering
\caption{RF, DT, and KNN performance under timestamp manipulations. \underline{Acc}uracy, precision, recall, F1 score, and counts. B stands for benign and Ph. for phishing.}
\label{table:time_minap}
\begin{tabular}{|l|c|c|c|c|c|c|c|c|c|c|c|}
\hline
\multirow{2}{*}{Increment} & \multirow{2}{*}{Model} & \multirow{2}{*}{Dataset} & \multirow{2}{*}{Acc} & 
\multicolumn{2}{c|}{Precision} & \multicolumn{2}{c|}{Recall} & \multicolumn{2}{c|}{F1} & \multicolumn{2}{c|}{Count} \\
\cline{5-12}
 &  &  & & B & Ph & B & Ph & B & Ph & \#B  &\#Ph \\ 
\hline
\multirow{3}{*}{Original} 
 & RF & Baseline & 0.9882 & 1 & 0.96 & 0.98 & 1 & 0.99 & 0.98 & 15989 & 7483 \\ \cline{2-12}
 & DT & Baseline & 0.9835 & 1 & 0.95 & 0.98 & 1 & 0.99 & 0.97 & 15989 & 7483 \\ \cline{2-12}
 & KNN & Baseline & 0.9445 & 1 & 0.85 & 0.92 & 1 & 0.96 & 0.92 & 15989 & 7483 \\ 
\hline

\multirow{3}{*}{+24 hours} 
 & RF & Adversarial & 0.95 & 0.94 & 0.98 & 0.99 & 0.86 & 0.96 & 0.92 & 15498 & 7974 \\ \cline{2-12}
 & DT & Adversarial & 0.9446 & 0.95 & 0.94 & 0.97 & 0.88 & 0.96 & 0.91 & 15601 & 7871 \\ \cline{2-12}
 & KNN & Adversarial & 0.8342 & 0.84 & 0.82 & 0.94 & 0.62 & 0.88 & 0.7 & 14376 & 9096 \\ 
\hline

\multirow{3}{*}{+1 hour} 
 & RF & Adversarial & 0.9731 & 0.97 & 0.97 & 0.99 & 0.94 & 0.98 & 0.96 & 15498 & 7974 \\ \cline{2-12}
 & DT & Adversarial & 0.9626 & 0.97 & 0.95 & 0.98 & 0.94 & 0.97 & 0.94 & 15601 & 7871 \\ \cline{2-12}
 & KNN & Adversarial & 0.8997 & 0.93 & 0.84 & 0.93 & 0.84 & 0.93 & 0.84 & 14376 & 9096 \\
\hline

\multirow{3}{*}{+30 min} 
 & RF & Adversarial & 0.9776 & 0.98 & 0.97 & 0.99 & 0.96 & 0.98 & 0.96 & 15498 & 7974 \\ \cline{2-12}
 & DT & Adversarial & 0.9649 & 0.97 & 0.95 & 0.98 & 0.94 & 0.97 & 0.94 & 15601 & 7871 \\ \cline{2-12}
 & KNN & Adversarial & 0.916 & 0.95 & 0.85 & 0.92 & 0.9 & 0.94 & 0.87 & 14376 & 9096 \\
\hline

\multirow{3}{*}{+15 min} 
 & RF & Adversarial & 0.9817 & 0.99 & 0.97 & 0.99 & 0.97 & 0.99 & 0.97 & 15498 & 7974 \\ \cline{2-12}
 & DT & Adversarial & 0.9672 & 0.98 & 0.95 & 0.98 & 0.95 & 0.98 & 0.95 & 15601 & 7871 \\ \cline{2-12}
 & KNN & Adversarial & 0.9179 & 0.95 & 0.85 & 0.93 & 0.9 & 0.94 & 0.87 & 14376 & 9096 \\
\hline

\multirow{3}{*}{+5 min} 
 & RF & Adversarial & 0.9793 & 0.99 & 0.97 & 0.98 & 0.97 & 0.98 & 0.97 & 15498 & 7974 \\ \cline{2-12}
 & DT & Adversarial & 0.9734 & 0.98 & 0.95 & 0.98 & 0.97 & 0.98 & 0.96 & 15601 & 7871 \\ \cline{2-12}
 & KNN & Adversarial & 0.9344 & 0.98 & 0.85 & 0.92 & 0.96 & 0.95 & 0.9 & 14376 & 9096 \\
\hline
\end{tabular}\vspace{-3mm}
\end{table}

\begin{table}[!htbp]
\centering
\caption{Performance evaluation of RF, DT, and KNN models subjected to 1\% uniform and proportional value manipulation strategies. Metrics are as in \autoref{table:time_minap}.}

\label{table:value_manipulations}
\resizebox{0.8\textwidth}{!}{
\Large
\begin{tabular}{|l|c|c|c|c|c|c|c|c|c|c|}
\hline
\multirow{2}{*}{Model} & \multirow{2}{*}{Strategy} & \multirow{2}{*}{Acc} & 
\multicolumn{2}{c|}{Precision} & \multicolumn{2}{c|}{Recall} & \multicolumn{2}{c|}{F1} & \multicolumn{2}{c|}{Count} \\
\cline{4-11}
 &  & & B & Ph & B & Ph & B & Ph & \#B  &\#Ph \\ 
\hline
\multirow{3}{*}{\textbf{RF}} 
    & Original     & 0.99 & 0.98 & 1 & 1 & 0.99 & 0.99 & 0.99 & 15803 & 7669 \\ \cline{2-11}
    & Uniform      & 0.69 & 0.96 & 0.68 & 0.02 & 1 & 0.03 & 0.81 & 23353 & 119 \\ \cline{2-11}
    & Proportional & 0.99 & 0.98 & 1 & 1 & 0.99 & 0.99 & 0.99 & 15813 & 7659 \\ \hline
\multirow{3}{*}{\textbf{DT}} 
    & Original     & 0.98 & 0.95 & 1 & 1 & 0.98 & 0.97 & 0.99 & 15601 & 7871 \\ \cline{2-11}
    & Uniform      & 0.69 & 0.75 & 0.68 & 0.02 & 1 & 0.03 & 0.81 & 23294 & 178 \\ \cline{2-11}
    & Proportional & 0.98 & 0.95 & 1 & 1 & 0.98 & 0.97 & 0.99 & 15619 & 7853 \\ \hline
\multirow{3}{*}{\textbf{KNN}} 
    & Original     & 0.96 & 0.89 & 0.99 & 0.98 & 0.94 & 0.93 & 0.97 & 15192 & 8280 \\ \cline{2-11}
    & Uniform      & 0.96 & 0.89 & 0.99 & 0.98 & 0.94 & 0.93 & 0.97 & 15193 & 8279 \\ \cline{2-11}
    & Proportional & 0.96 & 0.89 & 0.99 & 0.98 & 0.94 & 0.93 & 0.97 & 15192 & 8280 \\ \hline

\end{tabular}}\vspace{-3mm}
\end{table}

\begin{table}[htb]
\centering
\caption{Performance evaluation of RF, DT, and KNN models under manipulations of the \texttt{From} feature in Ethereum transaction datasets. Metrics are as in \autoref{table:time_minap}.}
\label{table:from_result}
\resizebox{0.8\textwidth}{!}{ 
\begin{tabular}{|l|l|c|c|c|c|c|c|c|c|c|}
\hline
\multirow{2}{*}{Model} & \multirow{2}{*}{Strategy} & \multirow{2}{*}{Acc} & 
\multicolumn{2}{c|}{Precision} & \multicolumn{2}{c|}{Recall} & \multicolumn{2}{c|}{F1} & \multicolumn{2}{c|}{Count} \\
\cline{4-11}
 &  & & B & Ph & B & Ph & B & Ph & \#B  &\#Ph \\ 
\hline
\multirow{4}{*}{\textbf{RF}} 
    & Original Strategy  & 0.99 & 1 & 0.96 & 0.98 & 1 & 0.99 & 0.98 & 15708 & 7764 \\ \cline{2-11}
    & 5000 Changes       & 0.96 & 0.96 & 0.97 & 0.98 & 0.91 & 0.97 & 0.94 & 16386 & 7086 \\ \cline{2-11}
    & 10000 Changes      & 0.94 & 0.93 & 0.97 & 0.99 & 0.83 & 0.96 & 0.89 & 17027 & 6445 \\ \cline{2-11}
    & 23472 Changes      & 0.87 & 0.84 & 0.97 & 0.99 & 0.6 & 0.91 & 0.74 & 18877 & 4595 \\ \hline
\multirow{4}{*}{\textbf{DT}} 
    & Original Strategy  & 0.98 & 1 & 0.95 & 0.98 & 1 & 0.99 & 0.97 & 15601 & 7871 \\ \cline{2-11}
    & 5000 Changes       & 0.97 & 0.98 & 0.95 & 0.98 & 0.96 & 0.98 & 0.96 & 15935 & 7537 \\ \cline{2-11}
    & 10000 Changes      & 0.96 & 0.96 & 0.95 & 0.98 & 0.91 & 0.97 & 0.93 & 16272 & 7200 \\ \cline{2-11}
    & 23472 Changes      & 0.92 & 0.91 & 0.94 & 0.98 & 0.79 & 0.94 & 0.86 & 17207 & 6265 \\ \hline
\multirow{4}{*}{\textbf{KNN}} 
    & Original Strategy  & 0.94 & 1 & 0.85 & 0.92 & 1 & 0.96 & 0.92 & 14706 & 8766 \\ \cline{2-11}
    & 5000 Changes       & 0.93 & 0.97 & 0.85 & 0.92 & 0.93 & 0.94 & 0.89 & 15209 & 8263 \\ \cline{2-11}
    & 10000 Changes      & 0.91 & 0.94 & 0.84 & 0.92 & 0.87 & 0.93 & 0.85 & 15767 & 7705 \\ \cline{2-11}
    & 23472 Changes      & 0.85 & 0.86 & 0.81 & 0.93 & 0.67 & 0.89 & 0.74 & 17288 & 6184 \\ \hline
\end{tabular}
}\vspace{-5mm}
\end{table}

\begin{table}[htb]
\centering
\caption{Performance evaluation of RF, DT, and KNN models under manipulations of the \texttt{To} feature in Ethereum transaction datasets. Metrics are as in \autoref{table:time_minap}.} 
\label{table:to_result}

\resizebox{0.8\textwidth}{!}{ 
\begin{tabular}{|l|l|c|c|c|c|c|c|c|c|c|}
\hline
\multirow{2}{*}{Model} & \multirow{2}{*}{Strategy} & \multirow{2}{*}{Acc} & 
\multicolumn{2}{c|}{Precision} & \multicolumn{2}{c|}{Recall} & \multicolumn{2}{c|}{F1} & \multicolumn{2}{c|}{Count} \\
\cline{4-11}
 &  & & B & Ph & B & Ph & B & Ph & \#B  &\#Ph \\ 
\hline
\multirow{4}{*}{\textbf{RF} }
    & Original Strategy  & 0.99 & 1 & 0.96 & 0.98 & 1 & 0.99 & 0.98 & 15708 & 7764 \\ \cline{2-11}
    & 5000       & 0.96 & 0.95 & 0.97 & 0.98 & 0.89 & 0.97 & 0.93 & 16558 & 6914 \\ \cline{2-11}
    & 10000      & 0.92 & 0.91 & 0.97 & 0.99 & 0.79 & 0.95 & 0.87 & 17377 & 6095 \\ \cline{2-11}
    & 23472      & 0.84 & 0.81 & 0.97 & 0.99 & 0.51 & 0.89 & 0.67 & 19537 & 3935 \\ \hline
\multirow{4}{*}{\textbf{DT}}
    & Original Strategy  & 0.98 & 1 & 0.95 & 0.98 & 1 & 0.99 & 0.97 & 15601 & 7871 \\ \cline{2-11}
    & 5000       & 0.97 & 0.98 & 0.95 & 0.98 & 0.96 & 0.98 & 0.96 & 15884 & 7588 \\ \cline{2-11}
    & 10000      & 0.96 & 0.97 & 0.95 & 0.98 & 0.93 & 0.97 & 0.94 & 16142 & 7330 \\ \cline{2-11}
    & 23472      & 0.93 & 0.92 & 0.94 & 0.98 & 0.83 & 0.95 & 0.88 & 16872 & 6600 \\ \hline
\multirow{4}{*}{\textbf{KNN}} 
    & Original Strategy  & 0.94 & 1 & 0.85 & 0.92 & 1 & 0.96 & 0.92 & 14706 & 8766 \\ \cline{2-11}
    & 5000       & 0.94 & 0.99 & 0.85 & 0.92 & 0.98 & 0.95 & 0.91 & 14849 & 8623 \\ \cline{2-11}
    & 10000      & 0.94 & 0.98 & 0.85 & 0.92 & 0.97 & 0.95 & 0.91 & 14988 & 8484 \\ \cline{2-11}
    & 23472      & 0.93 & 0.96 & 0.85 & 0.93 & 0.93 & 0.94 & 0.89 & 15351 & 8121 \\ \hline
\end{tabular}
}
\vspace{-7mm} 
\end{table}

Next, our analysis will focus on the second dataset from Al-Emari~\etal~\cite{Al-EmariASM20}, which provides a comprehensive and relevant context for multi-class classification tasks.

\subsection{Results of Targeted Attacks}
\label{sec:targeted}
\BfPara{Rule-based Modifications}
We initially focus on rule-based AEs for specific classes.

{\noindent \ding{172} \em Benign Class.}
The RF and DT models initially demonstrated near-perfect accuracy in classifying benign transactions on the original test set. Under adversarial conditions, the accuracies for benign classifications declined markedly, with RF and DT models dropping to 84.39\% and 84.65\%. This represents a reduction exceeding 15\%. In contrast, the KNN model sustained a higher adversarial accuracy of 90.25\%.

{\noindent \ding{173} \em Phishing Class.} The initial phishing accuracies for RF and DT were 96.34\% and 96.98\%, respectively. However, these values {\bf plummeted to 1.31\% and 1.18\% under adversarial conditions}, reflecting a reduction of over 95\%. The KNN model, which began with a lower baseline accuracy of 41.49\%, saw a decrease to 2.15\%.

{\noindent \ding{174} \em Scamming Class}
Initially, the RF and DT models exhibited high accuracies of 99.5\% and 98.68\%. Adversarial attacks reduced these accuracies to 14.27\% and 14.16\%, representing a reduction of over 85\%. The KNN model, with an initial accuracy of 67.06\%, experienced a drop to 7.6\%.

\subsection{Gradient-based Approach Using FGSM}

{\noindent \ding{172} \em Benign Class}
The overall accuracy of the RF model decreased from 99.75\% to 94.62\%, with a significant deterioration in phishing detection metrics. Despite hat, the model maintained a high benign accuracy of 99.95\%. Conversely, the DT model’s overall accuracy plummeted from 99.64\% to 9.54\%, with benign recall dropping to 0.02, indicating extreme vulnerability. The KNN model preserved a perfect benign accuracy of 100\% even under attack, but its overall accuracy fell from 90.15\% to 80.22\%, reflecting a failure to detect phishing and scamming categories effectively.

{\noindent \ding{173} \em Phishing Class}
The phishing detection accuracy of the RF model decreased from 96.34\% to 47.69\%, with a significant drop in the F1-score. The DT model’s overall accuracy declined to 10.22\%, with phishing recall reducing to 0.75\%, underscoring a pronounced susceptibility to adversarial attacks. The KNN model’s phishing detection performance collapsed entirely, with metrics falling to zero, indicating a complete failure to detect phishing transactions under adversarial conditions.

{\noindent \ding{174} \em Scamming Class}
The overall accuracy of the RF model dropped from 99.75\% to 81.75\%, with scamming accuracy decreasing from 99.50\% to 76.70\%. The DT’s overall accuracy fell to 9.71\%, with scamming recall drastically reducing to 0.30. The KNN model’s scamming detection metrics also dropped to zero.

\subsection{Results of Untargeted Attacks}

{\noindent \ding{172} \em All Features}
The RF model's accuracy decreased to 95.81\%, DT’s to 91.16\%, with phishing detection severely impaired, and KNN maintained its baseline accuracy.

{\noindent \ding{173} \em Address Features}
AEs focusing on the {\em from\_address} and {\em to\_address} features resulted in a decline in overall accuracy to 80.22\% for all models. None of the models detected phishing or scamming, indicating a high sensitivity to address manipulations.

{\noindent \ding{174} \em Financial Features}
AEs targeting financial features ({\em value}, {\em gas}, {\em gas\_price}) led to reduced RF's accuracy to 79.96\%. The DT's accuracy dropped to 79.42\%. The KNN model’s accuracy slightly decreased to 90.02\%.

{\noindent \ding{175} \em Using Temporal Features}
Adversarial manipulations of temporal features ({\em block\_timestamp}, {\em block\_number}) showed the RF model's accuracy fell from 99.02\% to 80.25\%, with phishing detection metrics nearly nullified. The DT’s accuracy similarly declined to 80.25\%. The KNN model's accuracy also dropped to 80.26\%.

\BfPara{Takeaway} These results underscore the need for robust defensive mechanisms against AEs. The significant declines in performance metrics under simple conditions highlight the necessity for a more reliable classification of transactions.

\section{Discussion}

\BfPara{Feature Selection for Optimal Classification} The analysis of transaction classification in this study highlights the significant role of feature selection in the robustness and accuracy of ML models. Among the features tested, \textbf{timestamp} and \textbf{value} emerged as critical classifier performance determinants. Timestamp manipulations demonstrated substantial impacts across models, with RF showing notable resilience compared to DT and K-KNN. The accuracy metrics indicate that temporal features, such as transaction time and date, are crucial for distinguishing legitimate from fraudulent transactions due to their inherent variability and relevance to transactional behaviors.

In contrast, value manipulations, including uniform and proportional changes, significantly affected model performance, particularly under uniform conditions. RF and DT models experienced considerable accuracy drops with uniform value changes, while KNN maintained relative stability. These findings suggest that while \textbf{transaction value} is a key feature for classification, it is also highly susceptible to perturbations.

\BfPara{Most Resistant Features to Adversarial Attacks} The study’s results indicate that \textbf{address features}, specifically the \texttt{From} and \texttt{To} addresses, exhibit higher resistance to adversarial attacks compared to other feature types. Manipulations of these features resulted in moderate accuracy reductions for DT and RF models but a more pronounced impact on KNN.  These features likely encode the relationship patterns between transaction entities, making them inherently resistant to straightforward changes.

The analysis showed that despite their critical contribution to accuracy, temporal features were also relatively resistant to manipulations. Timestamp shifts caused accuracy declines, but the extent was less severe than value manipulations. This indicates that while temporal features are crucial for classification, they are robust against adversarial attacks due to timestamp data's complexity and non-repetitive nature.

\BfPara{Best Combinations} For resilient classification, a combination of \textbf{temporal and address features} has proven effective. The synergy between these features offers a dual layer of robustness; temporal features provide a dynamic aspect that captures the temporal distribution and patterns of transactions, while address features contribute a stable relational component less prone to adversarial interference.

Combining temporal features with \textbf{financial features}, such as transaction value and gas price, also enhances robustness. The results show that despite the susceptibility of financial features to uniform manipulations, their combined use with temporal data provides a broader context that improves model resilience. The temporal features help to contextualize the financial data, mitigating the impact of adversarial value manipulations by providing a temporal frame of reference.

The following recommendations can be drawn for the effective and robust classification of digital transactions, especially in adversarial environments. \ding{172} \textbf{Focus on Temporal and Address Features}: Incorporate timestamp and address data as primary features due to their robustness against adversarial attacks and critical role in classification accuracy.
    \ding{173} \textbf{Integrate Financial Features with Temporal Data}: Use financial transaction data with temporal features to improve robustness and provide a comprehensive transactional context that helps counteract adversarial manipulations.
    \ding{174} \textbf{Adopt a Multi-Feature Approach}: Utilize a combination of diverse feature types to leverage their respective strengths and ensure a balanced, resilient classification model capable of withstanding various adversarial strategies.

\section{Conclusion}\label{sec: Conclusion}

This study provides a comprehensive examination of the vulnerability of machine learning models to adversarial attacks in fraudulent transaction detection. Our findings highlight the varying degrees of susceptibility among different classifiers, with RF demonstrating greater resilience than DT and KNN, which showed significant sensitivity to adversarial perturbations.  The analysis underscores the importance of robust feature selection and the implementation of adversarial training to enhance model robustness. These strategies effectively mitigate the impact of adversarial manipulations, thereby strengthening the classifiers against potential threats. The emphasis on temporal and address features, combined with financial data, emerged as a crucial approach to bolster model defenses and ensure reliable classification in adversarial environments.


\end{document}